%% file: isgt_final.tex
\documentclass[journal]{IEEEtran}
\usepackage{graphicx}
\usepackage{siunitx}
\usepackage{amsmath, amssymb, amsfonts}
\usepackage{cite}
\usepackage{url}
\usepackage{comment}
\usepackage{subcaption}
\usepackage[dvipsnames]{xcolor}
\usepackage[acronym]{glossaries}

\makeglossaries
\newacronym{bess}{BESS}{Battery energy storage systems}
\newacronym{ems}{EMS}{Energy Management Systems}
\newacronym{ecm}{ECM}{Equivalent Circuit Model}
\newacronym{der}{DER}{Distributed Energy Resources}
\newacronym{mpc}{MPC}{Model Predictive Control}
\newacronym{drl}{DRL}{Deep Reinforcement Learning}
\newacronym{rl}{RL}{Reinforcement Learning}
\newacronym{il}{IL}{Imitation Learning}
\newacronym{bc}{BC}{Behavior Cloning}
\newacronym{dagger}{DAgger}{Dataset Aggregation}
\newacronym{pv}{PV}{Photovoltaic}
\newacronym{lp}{LP}{Linear Programming}
\newacronym{ppo}{PPO}{Proximal Policy Optimization}
\newacronym{nlp}{NLP}{Non-Linear Programming}
\newacronym{milp}{MILP}{Mixed-Integer Linear Programming}
\newacronym{gail}{GAIL}{Generative Adversarial Imitation Learning}
\newacronym{airl}{AIRL}{Adversarial Inverse Reinforcement Learning}
\newacronym{dp}{DP}{Dynamic Programming}
\newacronym{tou}{ToU}{Time of Use}
\newacronym{soc}{SOC}{State of Charge}
\newacronym{dam}{DAM}{Day-Ahead Market}

\begin{document}

\title{Price Aware Power Split Control in Heterogeneous Battery Storage Systems}



\author{
    Sheng~Yin$^*$\textsuperscript{1,2},
    Vivek Teja~Tanjavooru$^*$\textsuperscript{1,2},
    Thomas~Hamacher\textsuperscript{2},
    Christoph~Goebel\textsuperscript{2},
    Holger Hesse\textsuperscript{1}\\[0.1cm]
    \textsuperscript{1}Kempten University of Applied Sciences, Kempten, Germany \\
    \textsuperscript{2}Technical University of Munich, Munich, Germany \\

    \thanks{Sheng Yin and Vivek Tanjavooru contributed equally to this work.}%
    \thanks{(e-mail: sheng.yin@tum.de; vivek.tanjavooru@tum.de)}%
    \thanks{Manuscript submitted for review.}
}

\maketitle

\begin{abstract}
This paper presents a unified framework for the optimal scheduling of battery dispatch and internal power allocation in \gls{bess}. This novel approach integrates both market-based (price-aware) signals and physical system constraints to simultaneously optimize (1) external energy dispatch and (2) internal heterogeneity management of \gls{bess}, enhancing its operational economic value and performance. This work compares both model-based \gls{lp} and model-free \gls{rl} approaches for optimization under varying forecast assumptions, using a custom Gym-based simulation environment. The evaluation considers both long-term and short-term performance, focusing on economic savings, \gls{soc} and temperature balancing, and overall system efficiency. In summary, the long-term results show that the \gls{rl} approach achieved 10\% higher system efficiency compared to \gls{lp}, whereas the latter yielded 33\% greater cumulative savings. In terms of internal heterogeneity, the \gls{lp} approach resulted in lower mean \gls{soc} imbalance, while the \gls{rl} approach achieved better temperature balance between strings. This behavior is further examined in the short-term evaluation, which indicates that \gls{lp} delivers strong optimization under known and stable conditions, whereas \gls{rl} demonstrates higher adaptability in dynamic environments, offering potential advantages for real-time BESS control.
\end{abstract}


\section{Introduction}
\gls{bess} play a vital role in enabling the global shift to renewable energy by mitigating intermittency and offering temporal flexibility. Their economic and technical success relies heavily on effective operational strategies \cite{Divya2009}. Traditional \gls{ems} often separate high-level energy scheduling decisions, such as scheduling charge and discharge times, from low-level power split control, which involves allocating power among battery strings. This separation, while simplifying \gls{ems} algorithmic design, typically leads to sub-optimal overall performance and economic outcomes. 

In commercial and utility-scale systems, which typically use multi-string architectures, independently dispatching battery strings enhances performance, reliability, and reduces aging \cite{Tanjavooru.2025}. These benefits are even more critical in heterogeneous systems, where variations in capacity, chemistry, aging, or thermal behavior introduce complex optimization challenges best addressed through integrated approaches that align economic signals with physical constraints \cite{Patsios2016}.

The widespread adoption of dynamic electricity pricing mechanisms creates opportunities for \gls{bess} operators to leverage price arbitrage by charging during low-price periods and discharging during high-price periods \cite{He2016}. Several studies have explored price-aware battery scheduling strategies using mixed-integer linear programming and dynamic programming \cite{Ci2016 , Wei2017}. However, these approaches typically treat battery systems as single entities with uniform characteristics, overlooking heterogeneity among battery strings.

Power split control in heterogeneous \gls{bess} primarily focuses on balancing operational parameters like \gls{soc} and temperature, as imbalances can accelerate degradation, reduce efficiency, and pose safety risks \cite{Ci2016}. Methods such as model predictive control for \gls{soc} balancing \cite{liang2023model} and integrated approaches addressing thermal management \cite{Tanjavooru.2025} target these challenges. Still, most strategies are decoupled from system-level dispatch, which reacts only to price signals, missing economic optimization opportunities. 

Recent works addressing the gap between economic scheduling and physical constraint management include a two-stage optimization approach that considers price arbitrage and battery lifetime \cite{Xu2018}, and the integration of cycling aging constraints into price-based scheduling algorithms \cite{Wu2020}. These typically follow a hierarchical structure, where high-level economic decisions are later adjusted to meet physical constraints \cite{Collath.2022b}. Fully unified frameworks that simultaneously optimize both aspects remain rare.

\gls{bess} management methods are typically model-based (e.g., \gls{lp}) or learning-based (e.g., \gls{rl}). Model-based approaches handle constraints and forecasts well but rely on accurate models, while learning-based methods better handle uncertainty and complex dynamics. However, studies comparing these approaches for integrated, price-aware control in heterogeneous \gls{bess}  are scarce, leaving key research gaps. This paper addresses these gaps with a unified framework that jointly optimizes battery scheduling and power split under both economic and physical constraints. Key contributions include:


\begin{enumerate}
\item An open-source, single-stage optimization framework combining economic and physical objectives for heterogeneous multi-string \gls{bess}.
\item A comparison of model-based optimization vs. model-free reinforcement learning under varying forecast assumptions.
\item A customizable simulation platform with detailed electro-thermal battery models.
\item Quantitative evaluation using metrics such as economic savings, \gls{soc} and temperature imbalance, and system efficiency.
\end{enumerate}


\section{Methodology}
\begin{figure}[h]
    \centering
    \includegraphics[width=1\columnwidth]{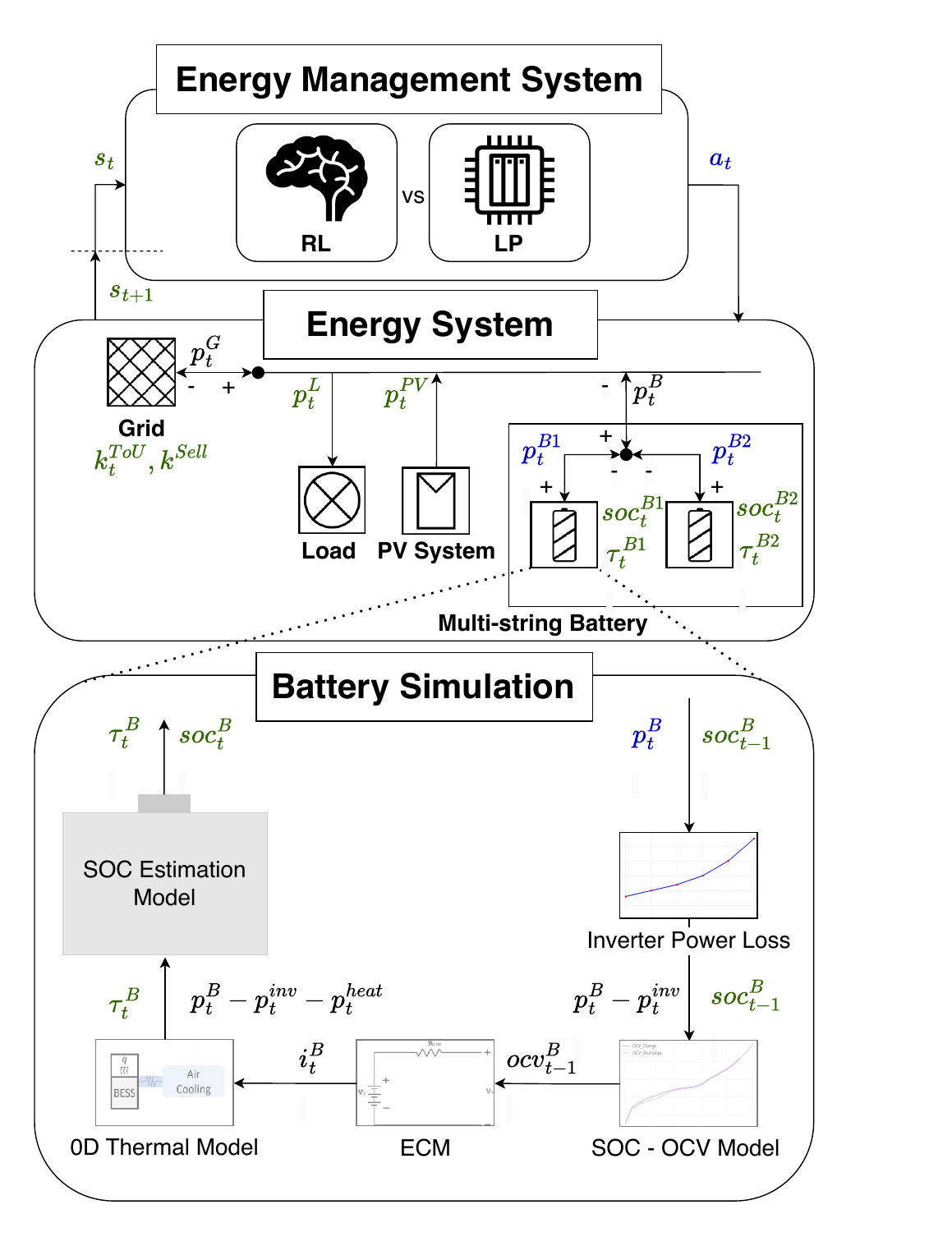} 
    \caption{Integrated EMS framework comprising the energy system of a multi-string \gls{bess} (here: two strings, A and B) with detailed battery simulation presenting control actions (blue) and observation states (green).}
    \label{fig:framework}
\end{figure}

This work presents an integrated framework for an optimal power scheduling strategy for a heterogeneous multi-string \gls{bess}, as shown in Figure \ref{fig:framework}. The \gls{bess} operates in a grid-connected industrial site (Section \ref{sec:system-def}). The \gls{bess} power schedule is determined by an \gls{ems}, where multiple methods are implemented and benchmarked (Section \ref{sec:control-approach}).

\subsection{Use Case Definition}
\label{sec:system-def}
The energy system we simulate comprises a User's Consumption (Load), a PV System, and a heterogeneous multi-string \gls{bess}, all connected to the power grid. The system operates under a time-of-use consumption tariff with electricity price $k^{\text{ToU}}_t$ at time $t$ and a static selling tariff with price $k^{\text{Sell}}$. At each time step t, the energy system is controlled by EMS's action \textbf{a} (shaded in blue) derived based on the system state \textbf{s} (shaded in green).

The objective is minimizing the total electricity cost of the energy system while ensuring optimal power split control by balancing both the \gls{soc} and the temperature of the battery strings, over the total time period $T=[t_1, t_2, t_3, \dots]$.

At each time step t, the energy cost is calculated as:
\begin{equation} 
\begin{aligned}
cost_t = \begin{cases}
p^{G}_t \cdot \Delta t \cdot k_{t}^{ToU} \cdot (1+ k), & \text{for $p^{G}_t  \geq 0$ },\\
p^{G}_t \cdot \Delta t \cdot k^{Sell} \cdot (1-k) , &\text{otherwise}.
\end{cases}
\end{aligned}
\label{eq:cost_policy}
\end{equation}
where $k$ is the tax ratio for energy trading and the grid power is calculated according to the power balance equation:
\begin{equation}
\label{eq:powerbalance}
p^{G}_t = p^{L}_t - p^{PV}_t + \sum_{m \in M}p^{B[m]}_t
\end{equation}
with $p^{L}_t$ denoting the load, $p^{PV}_t$ the PV generation, and $p^{B[m]}_t$ the battery power of unit $m$.

Assuming no battery usage, i.e., $\sum_{m \in M} p^{B[m]}_t = 0$, the resulting grid power $\Bar{p}^{G}_t$ and the corresponding cost without battery, denoted as $\overline{\text{cost}}_t$, can be computed using Equation~\eqref{eq:cost_policy}.

The battery temperature $\tau$ and \gls{soc} are calculated through a battery simulation model that integrates the electrical and thermal domains of a \gls{bess} through a cohesive set of interconnected models. It incorporates a nonlinear inverter power loss model that calculates inverter power loss, \(p^{\text{inv}}_{\text{t}}\),  based on throughput power, \(p^{\text{B}}_{\text{t}}\). The SOC–OCV model represents the electrochemical relationship between the battery’s \gls{soc} and its open circuit voltage and estimates the \(ocv^{\text{B}}_{\text{t}}\) of the battery using the \(soc^{\text{B}}_{\text{t}}\) from the battery's previous state - detailed formulation of this common battery modelling approach can be found \cite{Baccouche.2017b}. Additionally, an equivalent circuit model (ECM) is used to compute the cell level electrical current using the \(p^{\text{B}}_{\text{t}}\), \(ocv^{\text{B}}_{\text{t}}\) and internal resistance. These electrical models are coupled with a lumped mass thermal simulation model that calculates the heat generated by electrical losses and estimates the mean temperature of the battery system over time - an approach derived and described in detail in a previous work by the authors \cite{Tanjavooru.2025}. Finally, with all the power losses estimated during operation, the actual \gls{soc} is estimated using the Coulomb counting method \cite{Piller2001}. 

This electro-thermal coupling is critical for assessing system performance and supporting control strategies in this work that rely on energy and thermal constraints. The outputs from the controller such as the \gls{bess} power and predicted \gls{soc} act as the inputs for the simulation model to compute the mean temperature and actual \gls{soc}. 

\subsection{Control Approaches}
Two different control approaches are described and assessed in the following: 
\label{sec:control-approach}
\textbf{Linear programming (LP)}: 
This method adopts a rolling horizon-based deterministic linear programming technique to schedule and operate a \gls{bess}. It relies on forecasts of both load demand and PV generation, which may be based on either perfect foresight or persistence-based predictions~\cite{Moshovel.2015}. The optimization problem is formulated as a multi-objective framework, with the primary objective of minimizing the total operational cost, as expressed in Equation \eqref{eq:LP_obj}. In achieving this, the formulation simultaneously addresses secondary objectives, namely the balancing of the \gls{soc} and the uniformity of temperature $\tau$ across the battery strings. These secondary goals are critical for enhancing the performance, extending the lifespan, and ensuring the safety of the \gls{bess}. To ensure comparability among the different objectives, the coefficients ($x$,$y$,$z$) in the objective function are selected such that each individual objective is normalized. 

\begin{equation}
\label{eq:LP_obj}
\begin{aligned}
Objective = \min \Bigg(\sum_{t \in T}\left(x\cdot cost_t + y \cdot \Delta soc_t + z \cdot\Delta \tau_t\right) \Bigg)
\end{aligned}
\end{equation}

Subject to, for all $t \in T$ and $m \in M$:

\begin{equation}
p_t^{G,buy},\ p_t^{G,sell} \geq 0
\end{equation}

\begin{equation}
0 \leq p^{B[m],ch}_t,\ p^{B[m],dch}_t \leq p^{N}
\end{equation}

\begin{equation}
soc^{min} \leq soc^{B[m]}_t \leq soc^{max}
\end{equation}

\begin{equation}
soc^{B[m]}_{t} = soc^{B[m]}_{t-1} + \frac{\Delta t}{E^{N}} \cdot 
(p^{B[m],ch}_{t} \cdot \eta_{ch} - p^{B[m],dch}_{t}/\eta_{dch})
\end{equation}

\begin{equation}
soc_t^{mean} = \frac{1}{M} \sum_{m \in M} soc_t^{B[m]}
\label{soc_mean}
\end{equation}

\begin{equation}
\tau^{B[m]}_{t} = \tau^{B[m]}_{t-1} + \Delta t \cdot
\left(k1 \cdot p^{heat}_{t-1}
- k2 \cdot (\tau^{B[m]}_{t-1} - \tau_{\text{air}}) \right)
\end{equation}

\begin{equation}
\tau_t^{mean} = \frac{1}{M} \sum_{m \in M} \tau_t^{B[m]}
\label{T_mean}
\end{equation}

\begin{equation}
p^{G,buy}_t + p^{PV}_t + \sum_{m \in M}p^{B[m],dch}_t = p^{G,sell}_t + p^{L}_t + \sum_{m \in M}p^{B[m],ch}_t
\end{equation}

\begin{equation}
p^{B[m]}_t  = p^{B[m],ch}_t -  p^{B[m],dch}_t
\end{equation}

\begin{equation}
\begin{aligned}
cost_t = \left( p^{G,\text{buy}}_t \cdot k_{t}^{\text{ToU}} \cdot (1 + k) \cdot \Delta t \right) \\
\quad - \left( p^{G,\text{sell}}_t \cdot k_{t}^{\text{Sell}} \cdot (1 - k) \cdot \Delta t \right)
\end{aligned}
\end{equation}

\begin{equation}
\Delta soc_t = \sum_{m \in M} \left|  soc_t^{mean} -  soc_t^{B[m]} \right|
\label{eq:soc_objective}
\end{equation}


\begin{equation}
\Delta \tau_t = \sum_{m \in M} \left|  \tau_t^{mean} -  \tau_t^{B[m]} \right|
\label{eq:t_objective}
\end{equation}

\textbf{Learning-Based approach (RL):}
As learning-based controllers have gained popularity in recent years, we also implement a DRL control method based on a state-of-the-art hybrid approach suggested by Yin et al. \cite{yin2025boosting}. This method combines optimization with machine learning by enabling \gls{ppo} policy learning through cloning from Linear Programming solutions to efficiently derive high-performing control policies. This approach is adapted to the use case at hand of multi-string \gls{bess} under dynamic pricing conditions.

\paragraph{BC-LP PPO Framework}
The approach consists of three sequential phases: expert demonstration, behavior cloning, and reinforcement learning. First, expert trajectories are generated by solving a rolling-horizon \gls{lp} optimization problem, assuming perfect forecasts for load, PV generation, and electricity prices. These optimal trajectories serve as expert demonstrations for pre-training the policy.

In the second phase, a neural network policy $\pi_{\theta}$ is trained to imitate the LP expert policy using supervised learning, forming a behavior-cloned policy $\tilde{\pi}$. This behavior cloning serves as an efficient warm-start for the third phase, in which the policy is further refined through \gls{ppo} by interacting with the simulation environment. In the end, the optimal policy obtained by this approach $\tilde{\pi}^*$ is taken as the learning-based controller and benchmarked with the LP-based controller.

\paragraph{State-Action Space and Reward}
At each time step $t$, the system state $s_t$ is defined as:
\begin{equation}
\label{eq:state}
s_t = [p_t^L, p_t^{PV}, soc^{B1}_{t-1}, soc^{B2}_{t-1}, \tau^{B1}_{t-1}, \tau^{B2}_{t-1}, k_t^{ToU}]
\end{equation}
The action $a_t$ corresponds to the battery power set points for two battery strings:
\begin{equation}
\label{eq:action}
a_t = [p_t^{B1},p_t^{B2}]
\end{equation}
The reward is defined according to Equation \eqref{eq:LP_obj}:
\begin{equation} 
\label{eq:reward}
r_t  = x \cdot (\overline{cost}_t - cost_t )+ y \cdot \Delta soc_t + z \cdot\Delta \tau_t
\end{equation}
Note that the energy cost is not directly used as reward due to its dependency on input profiles. Instead, cost reduction as reward encourages the agent to maximize economic gains while respecting system constraints.

\section{Experiment Setup}
\subsection{Data and Computational Setup}
We use a publicly available dataset (\textit{EMSx dataset} \cite{le2023emsx}), which provides 15-minute resolution data from industrial sites with paired PV generation and electrical load profiles. For our experiments, we selected dataset \textit{ID 4}, covering a continuous period of 2 years and 2 months. The battery system is scaled to 500~kWh and 125~kW, composed of two strings: 300 kWh / 75 kW and 200 kWh / 50 kW. For pricing, we apply a fixed feed-in tariff of 0.086 €/kWh and dynamic purchase rates based on scaled Day-Ahead Market prices (0.18–0.38 €/kWh). The presented open source Python-based simulation framework includes:
\begin{enumerate}
   \item A Gym-style environment coupling battery models with energy management logic
   \item Gurobi \cite{gurobi} for LP solutions using rolling-horizon forecasts
   \item Stable Baselines3 \cite{stable-baselines3} for reinforcement learning, and Imitation \cite{gleave2022imitation} for behavior cloning
   \item An ActorCriticPolicy with two hidden layers (64 neurons, ReLU activations)
\end{enumerate}

\subsection{Scenarios and Forecast Settings}
To evaluate the performance of the controllers used in this framework under consistent external conditions, all simulations are conducted using identical load and pricing profiles, with both perfect and persistent forecast models applied across the scenarios. As detailed in Table~\ref{table: controllers}, two distinct simulation scenarios are designed to assess and compare the controllers’ performance. The first scenario represents a long-term operational setting with a simulation period of 365 days. This scenario begins with both battery systems at a minimum \gls{soc} (soc$^1$ = 0.1, soc$^2$ = 0.1) and mean battery temperatures set to $T1=T2=25^\circ$. It is intended to evaluate the controllers' long term performance focusing on total savings, system homogeneity during operation and efficiency. In contrast, the second scenario focuses on short-term adaptability with a simulation duration of 7 days. It introduces higher and asymmetric initial \gls{soc} levels (soc$^1$ = 0.7, soc$^2$ = 0.3), along with a thermal gradient ($\tau^1$ = 35$^\circ$C, $\tau^2$ = 25$^\circ$C), to test responsiveness to more dynamic and imbalanced starting conditions. Across both scenarios, three approaches are compared: a linear programming controller with perfect foresight (LP$^p$), a linear programming controller relying on persistent forecast model (LP$^f$), and a reinforcement learning-based controller. Here, the training process is repeated 10 times. Their performances are evaluated in Scenario 1 under the label $RL$. The best-performing instance, denoted as $RL^*$, is further evaluated in Scenario 2.

\begin{table}[h!]
\caption{Definition of Experimental Setup: Scenario 1 – Long-Term with balanced initial conditions; Scenario 2 – Short-Term with unbalanced \gls{soc} and Temperature.}
\centering
\renewcommand{\arraystretch}{1.5}
\begin{tabular}{|c|c|c|c|}
\hline
\textbf{Scenario} & \textbf{Initial Conditions} & \textbf{Controller} & \textbf{Duration}\\
\hline
1 & 
  \begin{tabular}[c]{@{}l@{}}
    $soc^1$ = 0.1 \\
    $soc^2$ = 0.1 \\
    $\tau^1$ = 25 \\
    $\tau^2$ = 25
  \end{tabular}& 
  \begin{tabular}[c]{@{}l@{}}
    $ RL$ \\
    $ LP^p $\\
    $ LP^f$ \\
    \end{tabular} & 365 days\\
\cline{3-3}
\hline
2 &
  \begin{tabular}[c]{@{}l@{}}
    $soc^1$ = 0.7 \\
    $soc^2$ = 0.3 \\
    $\tau^1$ = 35 \\
    $\tau^2$ = 25
  \end{tabular}&
  \begin{tabular}[c]{@{}l@{}}
   $ RL^*$ \\
   $ LP^p $\\
   $ LP^f$ \\
    \end{tabular} & 7 days\\
\hline
\end{tabular}
\label{table: controllers}
\end{table}

To evaluate the performance of the controllers across simulation scenarios, four key quantitative metrics are considered that capture different operational aspects of the energy management system under uniform pricing, load, and forecast conditions.

\begin{itemize}
    \item \textbf{Savings~(€):} Measures the total cost reduction achieved by each controller. Higher savings indicate more effective scheduling and battery usage under dynamic pricing.
    
    \item \textbf{\(\Delta\)\gls{soc}:} Quantifies the average deviation of individual battery strings’ \gls{soc} from the system-wide mean \gls{soc} over time, as defined in Eq.~\ref{eq:soc_objective}.

    \item \textbf{\(\Delta\)\scalebox{1.5}{$\tau$} (°C):} Represents temperature deviation across strings to evaluate thermal imbalance, which is critical for assessing aging and safety risks (see Eq.~\ref{eq:t_objective}).

    \item \textbf{System Efficiency ($\eta$):} Assesses how efficiently the system converts and stores energy, defined as:
    
    \begin{equation}
     \small
    \eta = \left( 1 - \frac{\sum_{t \in T}\sum_{m \in M}\left(p_t^{loss[m]}\cdot \Delta t\right)} {\sum_{t \in T}\sum_{m \in M}\left(\left|p_t^{B[m]}\right|\cdot \Delta t\right)} \right) \times 100
    \label{eta}
    \end{equation}

    \begin{equation}
     \small
    p_t^{BESS[m]} =  {\sum_{m \in M}\left(p_t^{B[m]}\right)}
    \label{eta}
    \end{equation}
    
    \begin{equation}
     \small
    p_t^{loss[m]} = p_t^{inv[m]} + p_t^{heat[m]} 
    \label{power_loss}
    \end{equation}

    where:
    \begin{align*}
        & p_t^{loss[m]} \text{ is the total loss through battery string-m}.
    \end{align*}
\end{itemize}

\section{Results and Discussion}
This section presents and analyzes the performance of the proposed approaches across key evaluation metrics, comparing long-term and short-term control behavior.

\subsection{Long-term Performance}
\begin{figure}[h]
    \centering
    \includegraphics[width=1\columnwidth]{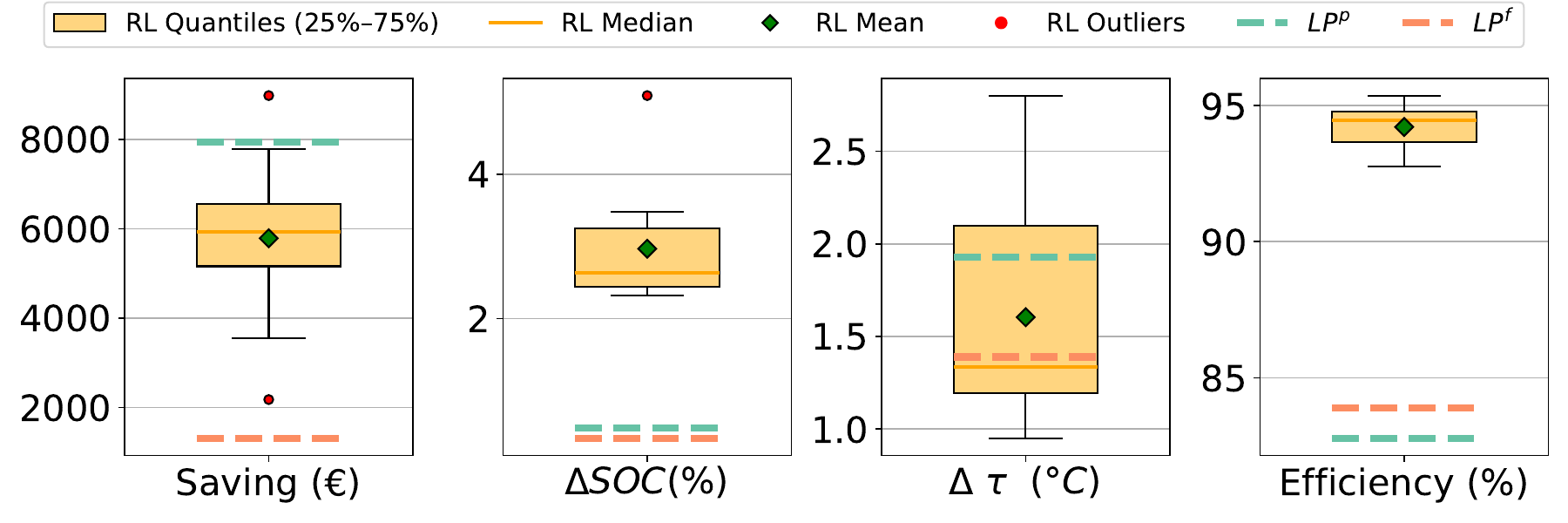} 
    \caption{Long-Term average performance assessment of RL and LP-based control.}
    \label{fig:metrics}
    
\end{figure}

Fig.~\ref{fig:metrics} presents the simulation results comparing the proposed models over 365 days across the four evaluation metrics averaged over time. The detailed results from the \textit{RL} solution including their 25$^{\text{th}}$ to 75$^{\text{th}}$ interquartile range, median, mean, and outliers are compared against the solutions from the deterministic optimization-based approaches, \textit{LP$^p$} and \textit{LP$^f$}. In terms of \textbf{Savings}, the benchmark \textit{LP$^p$} outperforms the baseline \textit{LP$^f$} and \textit{RL} solutions, achieving notably higher savings. For the average \textbf{$\Delta$SOC}, both \textit{LP}-based solutions exhibit lower deviation than the interquartile range of the \textit{RL} results, indicating more consistent control in homogenizing the SOC across strings. For average \textbf{$\Delta$\scalebox{1.5}{$\tau$}}, the \textit{RL} median yields lower thermal deviation than both LP solutions. The increased thermal spread observed for the   \textit{LP$^p$} case can be attributed to it's capability to exploiting potential savings at the cost of a more intensive battery operation (further details are provided in the next section focusing on the short term control performance).  In terms of \textbf{Efficiency ($\eta$)}, the RL model demonstrates higher than both the baseline \textit{LP$^f$} and benchmark \textit{LP$^p$} solutions, indicating better overall energy utilization.

In this scenario, the \textit{RL} outlier achieves even greater savings than the rest, possibly because the rolling horizon \textit{LP$^p$} solution, despite having perfect forecasts, is not globally optimal due to its limited foresight. Its multi-objective nature allows cost improvements through trade-offs with other metrics. Additionally, the simplified battery model used in the \textit{LP} optimizer reduces accuracy, creating further room for \textit{RL} to potentially outperform it.

\subsection{Control Performance}
\begin{figure}[h]
    \centering
    \begin{subfigure}[b]{0.9\columnwidth}
        \includegraphics[width=\linewidth]{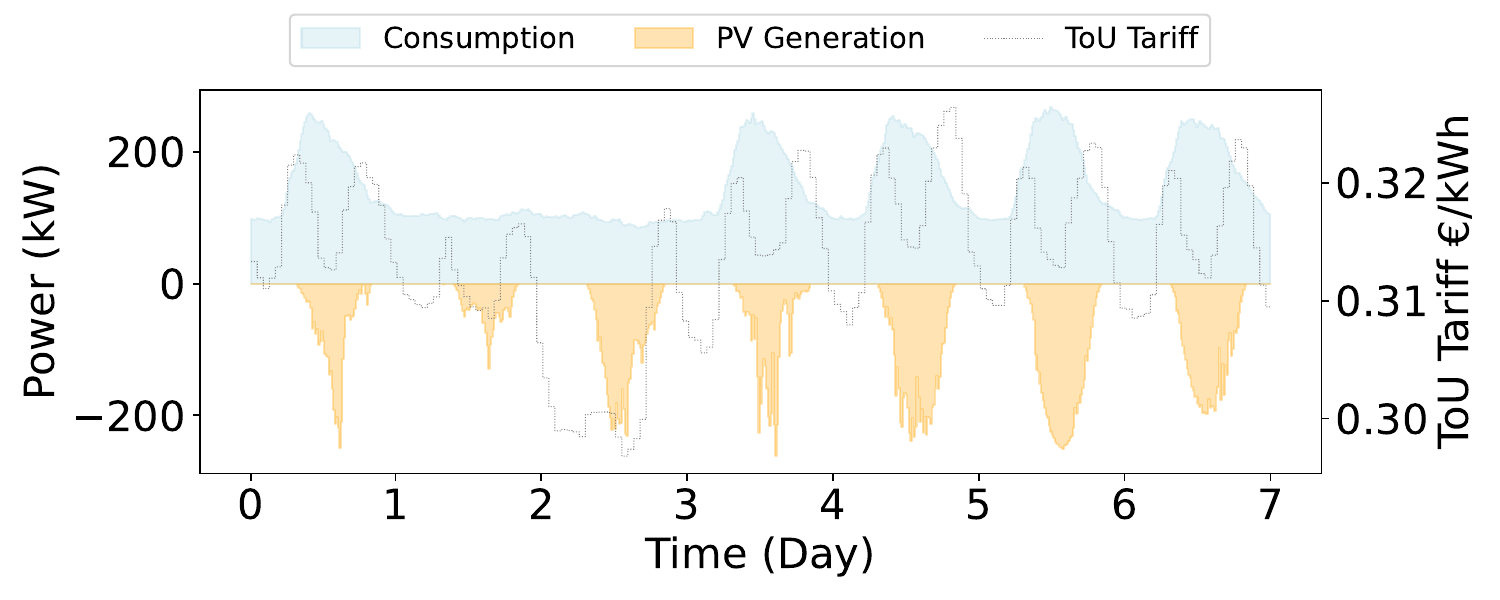}
        \label{fig:input_power}
    \end{subfigure}

    \caption{EMSx input dataset for ToU tariff.}
    \label{fig:input_data}
\end{figure}

In this section, the controllers developed in this work are evaluated for their short-term performance over a 7-day period, based on the evolution of the four metrics discussed in earlier sections. Fig.~\ref{fig:input_data} illustrates the corresponding 7-day input profile used for the \gls{bess} dispatch optimization problem, including PV generation, load/ consumption, and the Time-of-Use (ToU) tariff.

\begin{figure}[h]
    \centering
    
    \begin{subfigure}[b]{0.9\columnwidth}
        \includegraphics[width=\linewidth]{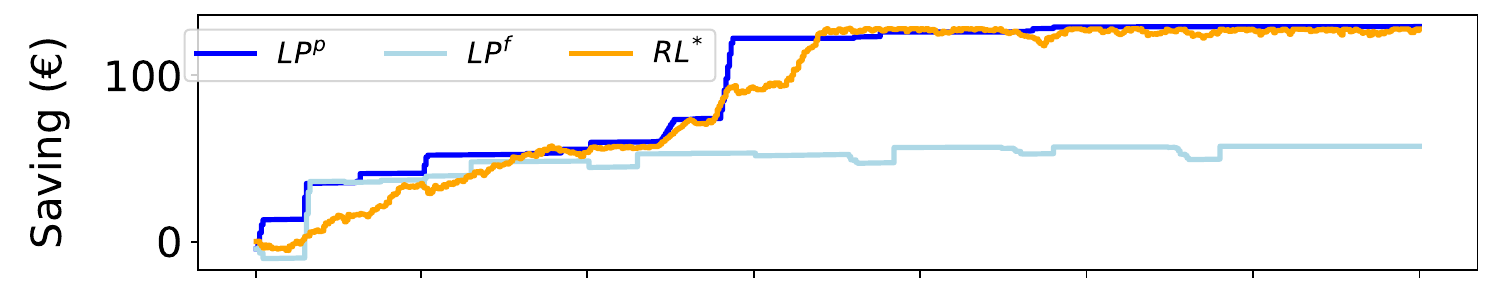}
        \label{fig:cumulative_Saving}
    \end{subfigure}
    
    \begin{subfigure}[b]{0.9\columnwidth}
        \includegraphics[width=\linewidth]{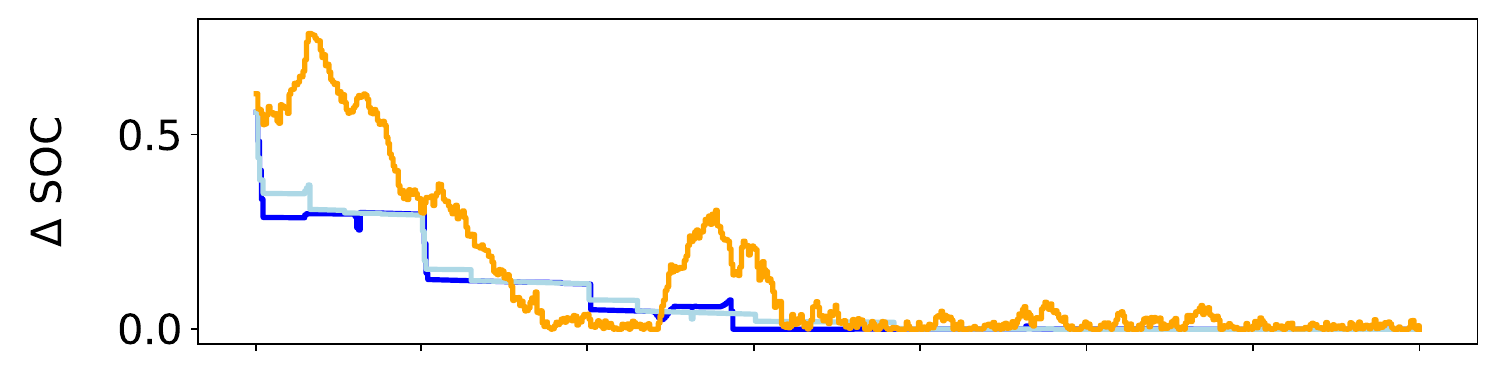}
        \label{fig:compare_soc}
    \end{subfigure}
    
    \begin{subfigure}[b]{0.9\columnwidth}
        \includegraphics[width=\linewidth]{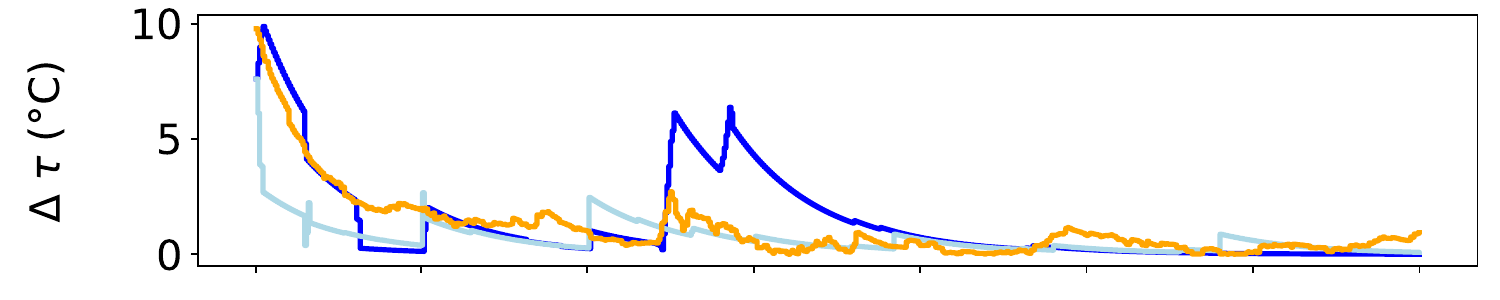}
        \label{fig:compare_t}
    \end{subfigure}
    
    \begin{subfigure}[b]{0.9\columnwidth}
        \includegraphics[width=\linewidth]{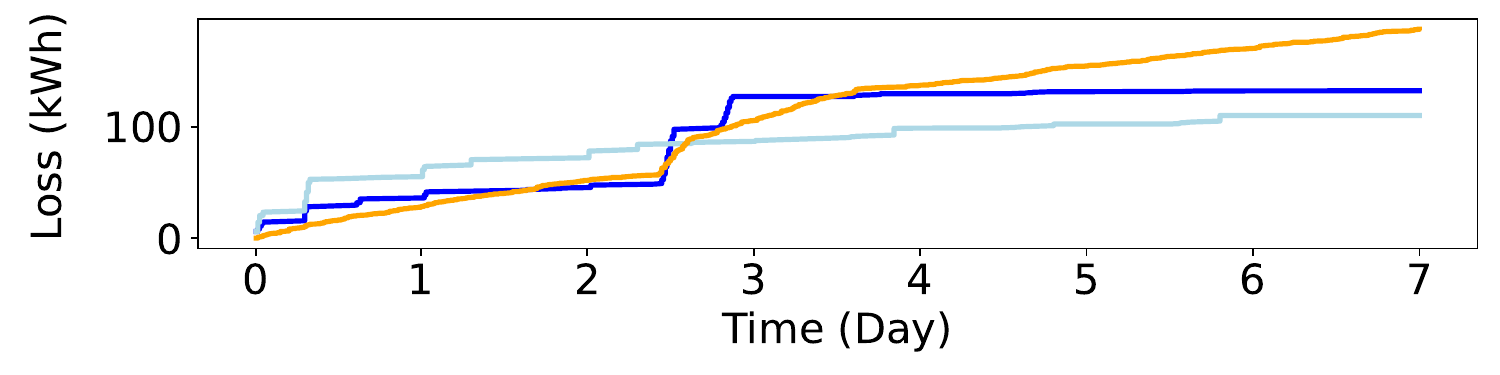}
        \label{fig:compare_t}
    \end{subfigure}
    
    \caption{Short-term control performance of RL and LP based controllers based on metrics evolution.}
    \label{fig:compare_all}
\end{figure}

The simulation results presented in Fig.~\ref{fig:compare_all} compare the best-performing RL model in terms of savings, \textit{RL$^*$}, against the LP-based solutions. The results suggest that the availability of a perfect forecast enables the \textit{LP$^p$} solution to outperform others in maximizing cost savings and homogenizing the battery strings with respect to SOC. However, by accurately forecasting a more active operational phase between the 2\textsuperscript{nd} and 3\textsuperscript{rd} days, it also concedes greater heterogeneity in temperature distribution across the strings. \textit{RL$^*$} performs better than the persistent forecast dependent \textit{LP$^f$} solution and on par with the \textit{LP$^p$} solution in all the four metrics but fails to capture the high operational activity which might lead to temperature heterogeneity. With the primary goal to generate maximum savings, the \textit{RL$^*$} model also engages the batteries with smaller throughput powers during the later part of the week, thereby incurring more inverter and thermal losses. This, in turn, led to higher overall energy losses for the \textit{RL$^*$} model compared to the LP solutions, as shown in the Loss (kWh) plot in Fig.~\ref{fig:compare_all}.

\subsection{Balance Performance}

\begin{figure}[h]
    \centering
    \begin{subfigure}[b]{1\columnwidth}
        \includegraphics[width=\linewidth]{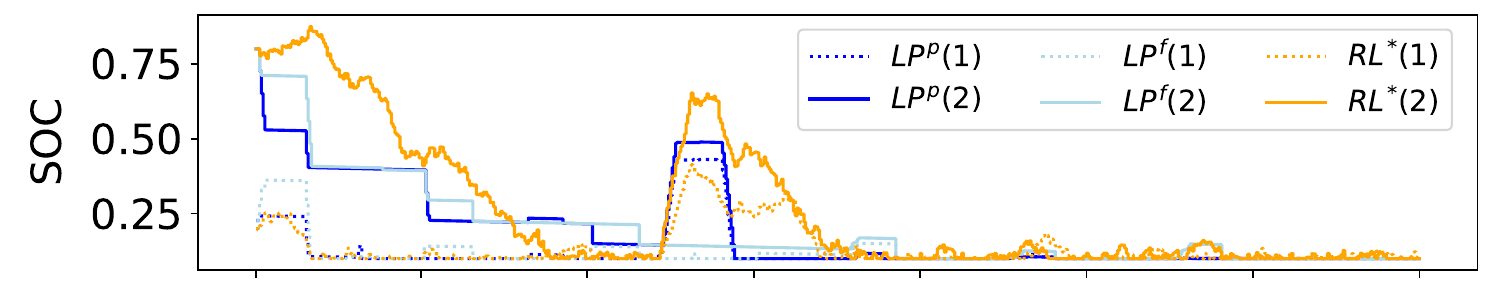}
        \label{fig:split_soc}
    \end{subfigure}
    
    \begin{subfigure}[b]{1\columnwidth}
        \includegraphics[width=\linewidth]{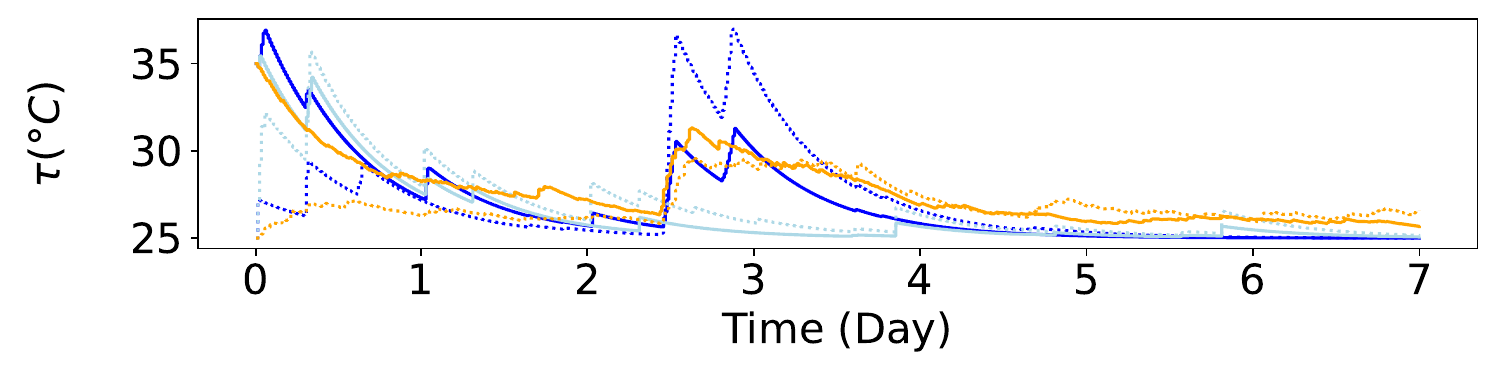}
        \label{fig:split_t}
    \end{subfigure}
    
    \caption{Comparison of system balancing capabilities of RL and LP based controllers.}
    \label{fig:split all}
\end{figure}

A deeper analysis of the secondary objectives, balancing key operational parameters such as SOC and temperature, is presented in Fig.~\ref{fig:split all}. The results show that \textit{RL$^*$} performs competitively with the benchmark \textit{LP$^p$} in achieving SOC and temperature uniformity. As previously discussed, the availability of a perfect forecast allows \textit{LP$^p$} to control battery states more accurately than \textit{RL$^*$}. However, under a more realistic scenario considering forecast uncertainty, \textit{RL$^*$} with learning-based adaptability outperforms \textit{LP$^f$} in controlling and estimating these critical parameters.

\section{Conclusion}
This paper presented a unified framework for simultaneously optimizing battery dispatch and power split control in heterogeneous BESS. Our comparison of model-based LP and model-free RL approaches revealed distinct strengths: LP achieved 33\% greater savings and better SOC balance under perfect forecasts, while RL demonstrated 10\% higher efficiency and superior temperature uniformity with greater adaptability to dynamic conditions. The findings highlight critical trade-offs: RL offers forecast-independent operation but requires extensive training; LP provides interpretable solutions for stable conditions but lacks flexibility with forecast uncertainty. Future research should explore hybrid approaches combining these strengths, improve forecasting methods, and implement higher-fidelity battery models to enhance real-world performance. This open-source framework establishes a foundation for integrated BESS management strategies that effectively address economic optimization with physical constraints.

\input{isgt_final.bbl}

\printglossaries

\end{document}